\documentstyle[psfig,subfig,apalike]{article}
\newcommand{\M}{{\cal M}}
\newcommand{\ie}{{\it i.e. }}
\newcommand{\eg}{{\it e.g. }}
\newcommand{\al}{{\it et al. }}
\newcommand{\be}{\begin{equation}}
\newcommand{\ee}{\end{equation}}
\newcommand{\sction}[1]{\section{#1 \hrulefill}}   

\title{Resampling Method For Unsupervised Estimation Of Cluster Validity}
\author{Erel Levine and Eytan Domany \\
Department of Physics of Complex Systems, \\
The Weizmann Institute of Science, 
Rehovot 76100, Israel}
\date{\today}
\begin{document}

\maketitle

\begin{abstract}
We introduce a method for validation of  
results obtained by clustering analysis of data. 
The method is based on resampling the available data. A figure
of merit that measures the stability of clustering solutions 
against resampling is introduced. 
Clusters which
are stable against resampling give rise to local maxima of this
figure of merit. This is presented first for a one-dimensional data set,
for which an analytic approximation for the figure of merit is derived
and compared with numerical measurements. Next, the applicability of 
the method is demonstrated for higher dimensional data, including 
gene microarray expression data.

\end{abstract}



\sction{Introduction}
Cluster analysis is an important  tool to investigate and interpret data.
Clustering techniques are the main tool used 
for exploratory data analysis, namely when one is
dealing with data about whose internal structure little or no
prior information is available. 
Cluster algorithms are expected to produce  partitions 
that reflect the internal structure of the data and identify ``natural" classes 
and hierarchies present in it. 

Throughout the years a wide variety of clustering algorithms 
have been proposed. Some
algorithms have their origins in graph theory, whereas others are based on 
statistical pattern recognition,
self-organization methods and more. More recently, several 
algorithms, which are rooted in statistical mechanics, have been introduced.

Comparing the relative merits of various methods is made difficult by the
fact that when applied to {\it the same} data set, 
different clustering algorithms 
often lead to {\it markedly different results}. In some cases such differences
are expected, since different algorithms make different (explicit or implicit)
assumptions about the  structure
of the data. If the particular set that is being studied consists, for example,
of several clouds of data point, with each cloud spherically distributed about
its center, methods that assume such structure (e.g. k-means), will work well.
On the other hand, if the data consist of a single non-spherical cluster, the
same algorithms will fare miserably, breaking it up into a hierarchy of
partitions.
Since for the cases of interest one does not know which assumptions are
satisfied by the data,
a researcher may run into severe
difficulties in interpretation of his results; by preferring one algorithm's
clusters over those of another, he may re-introduce his biases about the
underlying structure - precisely those biases, which one hoped to eliminate
by employing clustering techniques.
In addition, the differences in the sensitivity of 
different algorithms to noise,
which inherently exists in the data, also yield a major contribution
to the difference between their results. 

The ambiguity is made even more severe by the fact that even when one 
sticks exclusively to one's favorite algorithm, the results may depend strongly
on the values assigned to various parameters of the particular algorithm. 
For example, if there is a parameter which controls the resolution at
which the data are viewed, the algorithm produces a hierarchy of clusters (a
dendrogram) as a function of this parameter. One then has to decide which level
of the dendrogram reflects best the ``natural" classes present in the data?

Needless to say, one wishes to answer these questions in an unsupervised manner,
i.e. making use of nothing more than the available data itself.
Various methods and indicators, that come under the name ``cluster validation",
attempt to evaluate  the results of  cluster analysis in this manner
\cite{Jain88}. Numerous
studies suggest direct and indirect indices for evaluation of hard
clustering \cite{Jain88,Bock85}, probabilistic clustering \cite{Duda73}, and
fuzzy clustering
\cite{Windham82,Pal95} results. Hard clustering indices are often based on some
geometrical motivation to estimate how compact and well separated
the clusters are (e.g. Dunn's index \cite{Dunn74} and its generalizations
\cite{Bezdek95}); others are statistically motivated (e.g. comparing the
within cluster scattering with the between-cluster separation
\cite{Davies79}). Probabilistic and fuzzy indices are not considered
here. Indices proposed for these methods are based on likelihood-ratio tests
\cite{Duda73}, information-based criteria \cite{Cutler94} and more. 

Another approach to cluster validity includes some variant of
cross-validation \cite{Fukunaga90}. Such methods were introduced both
in the context of hard clustering \cite{Jain86} and fuzzy clustering
\cite{Smyth96}. The approach presented  here falls into this category.

In this paper we present a method to help select which clustering result is more
reliable. The method can be used to compare different algorithms, but it is most
suitable to identify, within the same algorithm, 
those partitions that can be attributed to the presence of noise.
In these cases a slight modification of the noise may alter the cluster
structure significantly. Our method controls and alters the noise by means of
{\it resampling} the original dataset.

In order to illustrate the problem we wish to address and its proposed 
solution, consider the following example. 
Say  a scientist is investigating mice, 
and comes to suspect that there are several types of them.
She therefore measures two features of the mice (such as weight and
shade), looks for clusters in this two-dimensional data set, and indeed
finds two clusters. She can therefore conclude that there are two
types of mice in her lab. Or can she?

\begin{figure}
 \centerline{
    \psfig{figure=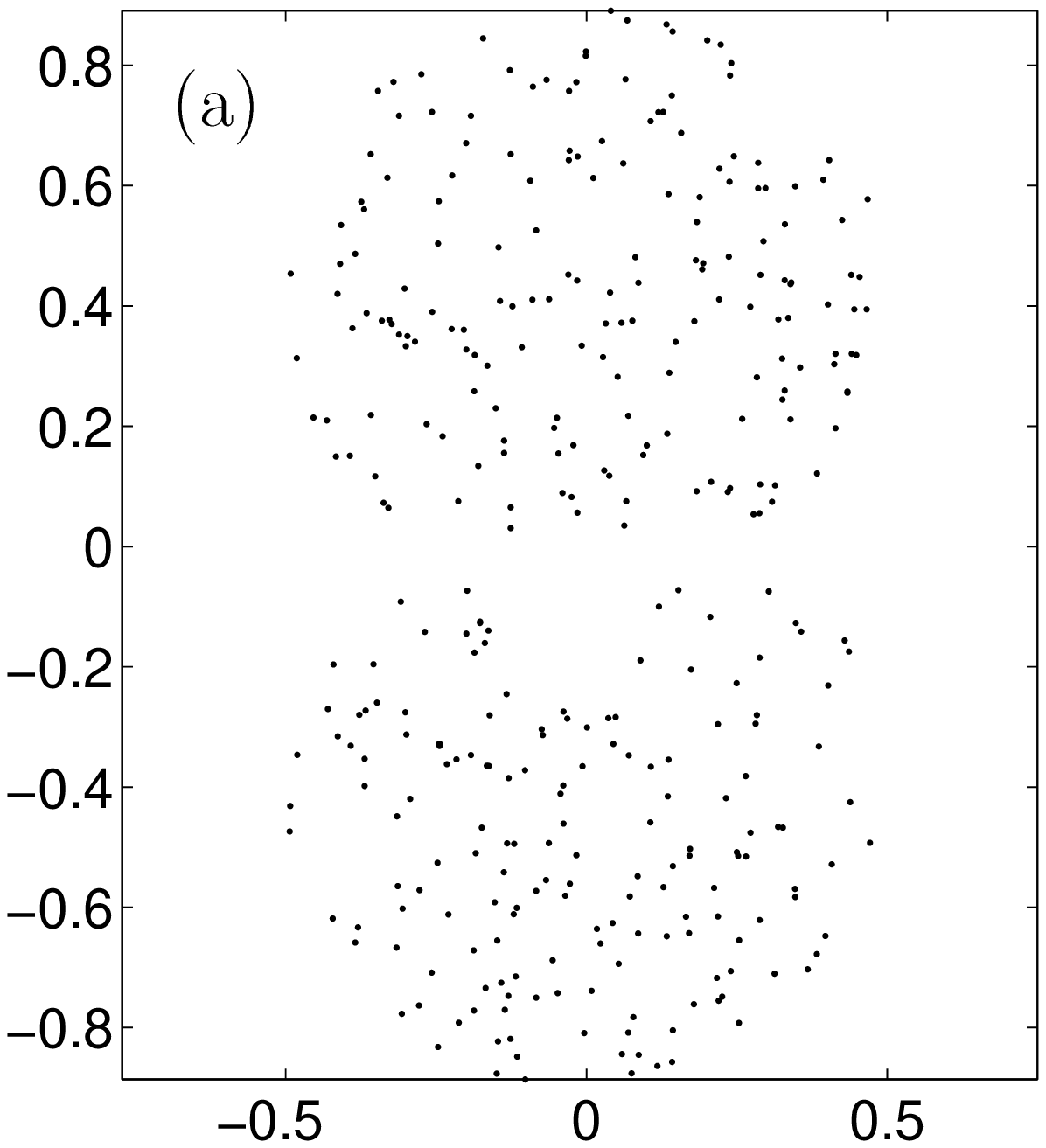,width=5.0cm} \hspace{1.0cm}
    \psfig{figure=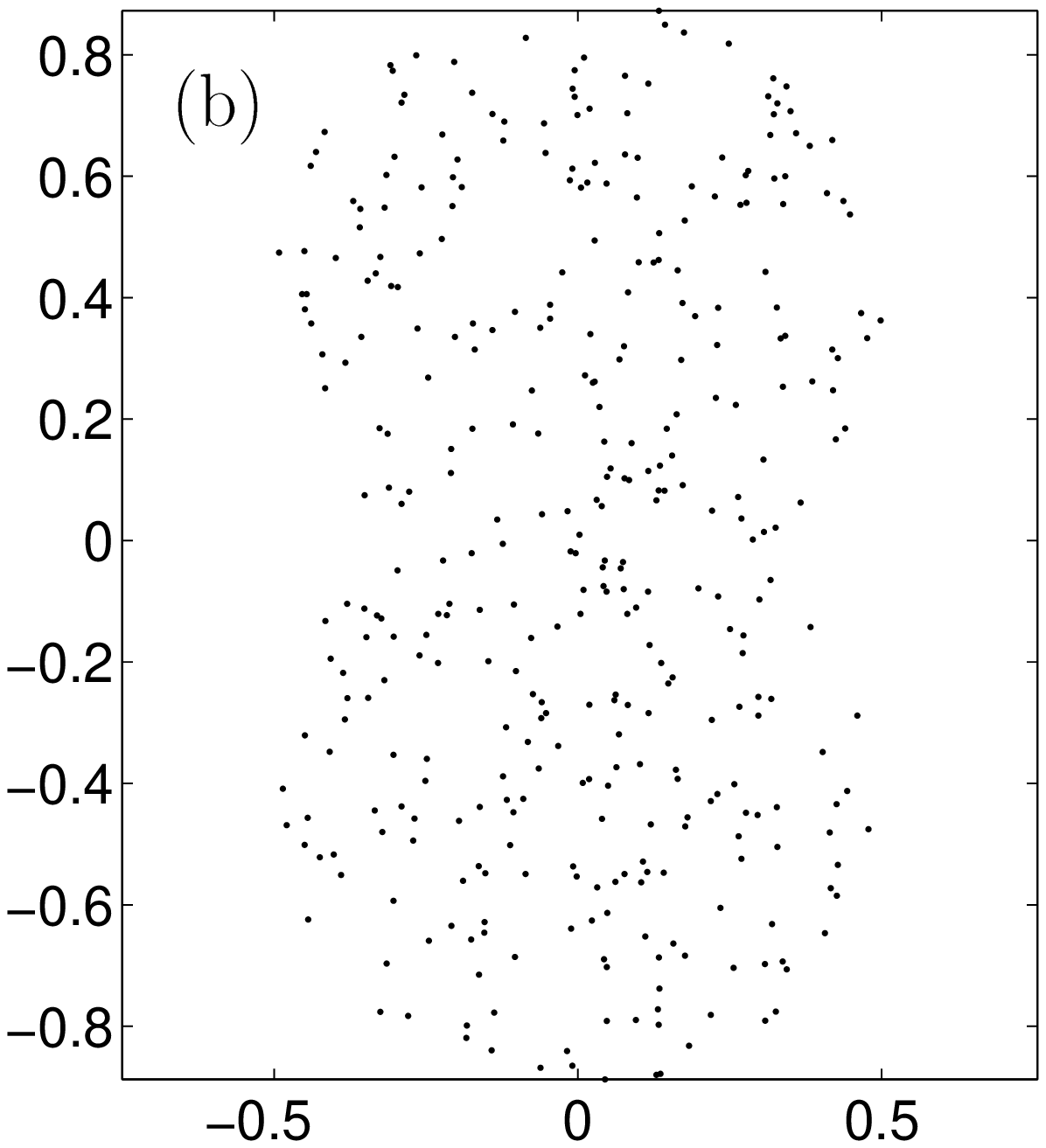,width=5.0cm}
  }
  \caption{Two samples, drawn from an 8-shaped uniform distribution.
           Sample (a) is somewhat non-typical. }
  \label{fig:example}
\end{figure}

Imagine that the data collected by the scientist can be represented by
the points on figure \ref{fig:example} (a). This data set was in fact
taken from  a shaped uniform distribution (with a relatively narrow ``neck" at 
the middle). Hence 
no partition really exists in the underlying
structure of the data and, unless one makes explicit assumptions about the shape 
of the data clouds, one should identify a single cluster.
The particular cluster algorithm used breaks, however, the
data into two clusters along the horizontal gap seen in 
figure \ref{fig:example} (a).
This gap happens to be the result of fluctuations in the data, or noise
in the sampling (or measurement) process. More typical data sets,
such as that of figure \ref{fig:example} (b), do not have such
gaps and are not broken into two clusters by the algorithm.

If more than a single sample had been available, it would have been safe to
assume that this particular gap would not have had appeared in most samples. The
partition into two clusters would, in that case, be easily identified as
unreliable. In most cases, however, only a single sample is available,
and resampling techniques are needed in order to generate several
ones. 

In this paper we propose a cluster validation method which is based
on resampling \cite{Good99,Efron93}: subsets of the data under
investigation are constructed randomly, and the cluster algorithm is
applied to each subset. The resampling scheme is introduced in Sec. 2,
and a figure of merit is proposed to identify
the stable clustering solutions, 
which are less likely to be the results of noise
or fluctuations. 
The proposed procedure is tested in Sec. 3
on a one-dimensional data set, for which
an analytical expression for the figure of merit is derived and compared
with the corresponding numerical results. In Sec. 4 we demonstrate the
applicability of our method to two artificial data sets (in $d=2$ dimensions)
and to real (very high dimensional) DNA microarray data.

\sction{The Resampling Scheme}
Let us denote the number of data points in the set to be studied by $N$.
Typically,  application
of any cluster algorithm necessitates choosing 
specific values for  some parameters.
The results yielded by the clustering algorithm may depend strongly on this choice. 
For example, some algorithms
(\eg the {\em c-shell} fuzzy-clustering \cite{Bezdek81} and the iso-data algorithm \cite{cover91}) 
take the expected number of clusters as part of their input.
Other algorithms (\eg Valley-Seeking \cite{Fukunaga90}) have the number
of neighbors of each point as an external parameter.

In particular, many algorithms 
have a parameter that controls the resolution at which clusters are identified.
In agglomerative clustering methods, for example, this parameter defines the
level of the resulting dendrogram at which the clustering solution is identified \cite{Jain88}.
For the K-nearest-neighbor algorithm \cite{Fukunaga90} a change in the number of
neighbors of each point, $K$, controls the resolution. 
For Deterministic Annealing and the Superparamagnetic Clustering
(SPC) algorithm this role is played by the {\it temperature T}.

As this control parameter is varied, the data-points get assigned to different
clusters, giving rise to a
 {\it hierarchy}. At the lowest resolution all $N$ points belong to one cluster,
 whereas at the other extreme one has $N$ clusters, of a single point in each.
 As the resolution parameter varies, clusters of 
data points break into sub-clusters which break further at a higher level. 
Sometimes the aim is to generate precisely such a dendrogram. In other cases
one would like to produce a single partitioning of the data, which captures a
particular important aspect. In such a case we wish to identify that value of
the resolution control parameter, at which the most reliable, natural 
clusters appear. In these situations the resolution parameter plays the role of
one (probably the most important) member of the family of parameters of 
the algorithm that needs to be fixed. Let us denote the full set of parameters
of our algorithm by $V$.

Any particular clustering solution can be presented
in the form of an  $N \times N$ {\it cluster connectivity matrix}  
${\cal T}_{ij}$, defined by
\begin{equation}
\label{eq:defT}
{\cal T}_{ij} = \left\{
\begin{array}{ll}
 1 & \mbox{points $i$ and $j$ belong to the same cluster}\\
 0 & \mbox{otherwise}
\end{array} \right.
\end{equation}
In order to validate this solution, we construct an ensemble
of $m$ such matrices, and make comparisons among them. This ensemble is 
created by 
constructing $m$ resamples of the original data set. A resample 
is obtained by selecting at random a
subset of size $fN$ of the data points. We call $0 \leq f \leq 1$ 
the {\em dilution factor}.

We apply to every one of these subsets the same clustering procedure, 
that was used on the full dataset, 
using the same set of
parameters $V$. This way we obtain
for each resample $\mu,~~\mu=1,...m,$ its own clustering results, summarized by
the $fN \times fN$ matrix ${\cal T}^{(\mu)}$. 

We define a {\it figure of merit} $\M (V)$ for the clustering procedure 
(and for the choice of parameters) 
that we used. The figure of merit is based on 
comparing the connectivity matrices of the resamples,
${\cal T}^{(\mu)} \;\;(\mu=1\ldots m)$, with the original matrix ${\cal T}$:
\begin{equation}
\label{eq:M}
\M (V) =  \ll \delta_{{\cal T}_{ij}, {\cal T}^{(\mu)}_{ij}} \gg _m \; ,
\end{equation}
The averaging implied by the  notation $\ll \cdot \gg _m$ is twofold.
First, for each resample $\mu$ we average over all those
pairs of points $ij$ which were neighbors\footnote{For 
various definition of neighbors see \cite{Fukunaga90}} 
in the original sample, and have both survived the
resampling. Second, this average value is averaged over all 
the $m$ different resamples. Clearly, $0 \leq \M \leq 1$, with $\M=1$ for
perfect score. 

The figure of merit $\M$ measures the extent to which the clustering assignments
obtained from the resamples agrees with that of the full sample.
An important assumption we have made implicitly in this procedure is that the
algorithm`s parameters are ``intensive", 
\ie their effect on the quality of the result is 
independent of the size of the data set. We can generalize our procedure
in several ways
to cases when this assumption does not hold\footnote{
For example, we can define our figure of merit on the basis of pairwise
comparisons of our resamples, and find an optimal set of parameters $V^*_1$
in the way explained below. Next, we look for
parameters $V^*_2$ for which clustering of the full sample yields closest 
results to those 
obtained for the resamples (clustered at  $V^*_1$).} \cite{Levine00}.

After calculating $\M (V)$, we have to decide whether we accept the clustering
result, obtained using a particular value of the clustering parameters,
or not. For very low and very high values of $\M$ the decision
may be easy, but for mid-range values we may need some additional information to
guide our decision. In such a case,
the best way to proceed is to change the values of the clustering parameters,
and go through the whole process once again.

Having done so for some set parameter choices $V$, we study the way 
$\M(V)$ varies as a function of $V$. Optimal
sets of parameters $V^*$ are  identified by locating the maxima of this 
function.
It should be noted, however, that some of these maxima are
trivial and should not be used. Examples that demonstrate this point are 
presented in the next section.

Our procedure can be summarized
in the following algorithmic form:
\begin{description}
\item[ Step 0.] Choose values for the parameters $V$ of the clustering algorithm.
\item[ Step 1.] Perform clustering analysis of the full data set.
\item[ Step 2.] Construct $m$ sub sets of the data set, by randomly
  selecting $fN$ of the $N$ original data-points. 
\item[ Step 3.] Perform clustering analysis for each sub set.
\item[ Step 4.] Based on the clustering results obtained in {\bf Steps
  1} and {\bf 3} calculate $\M(V)$, as defined in
  eq. (\ref{eq:M}).
\item[ Step 5.] Vary the parameters $V$ and identify stable clusters as 
those for which a local
  maximum of $\M$ is observed.
\end{description}

\begin{figure}[htbp] 
\begin{tabbing}
     \subfigure[]
     {
        \psfig{figure=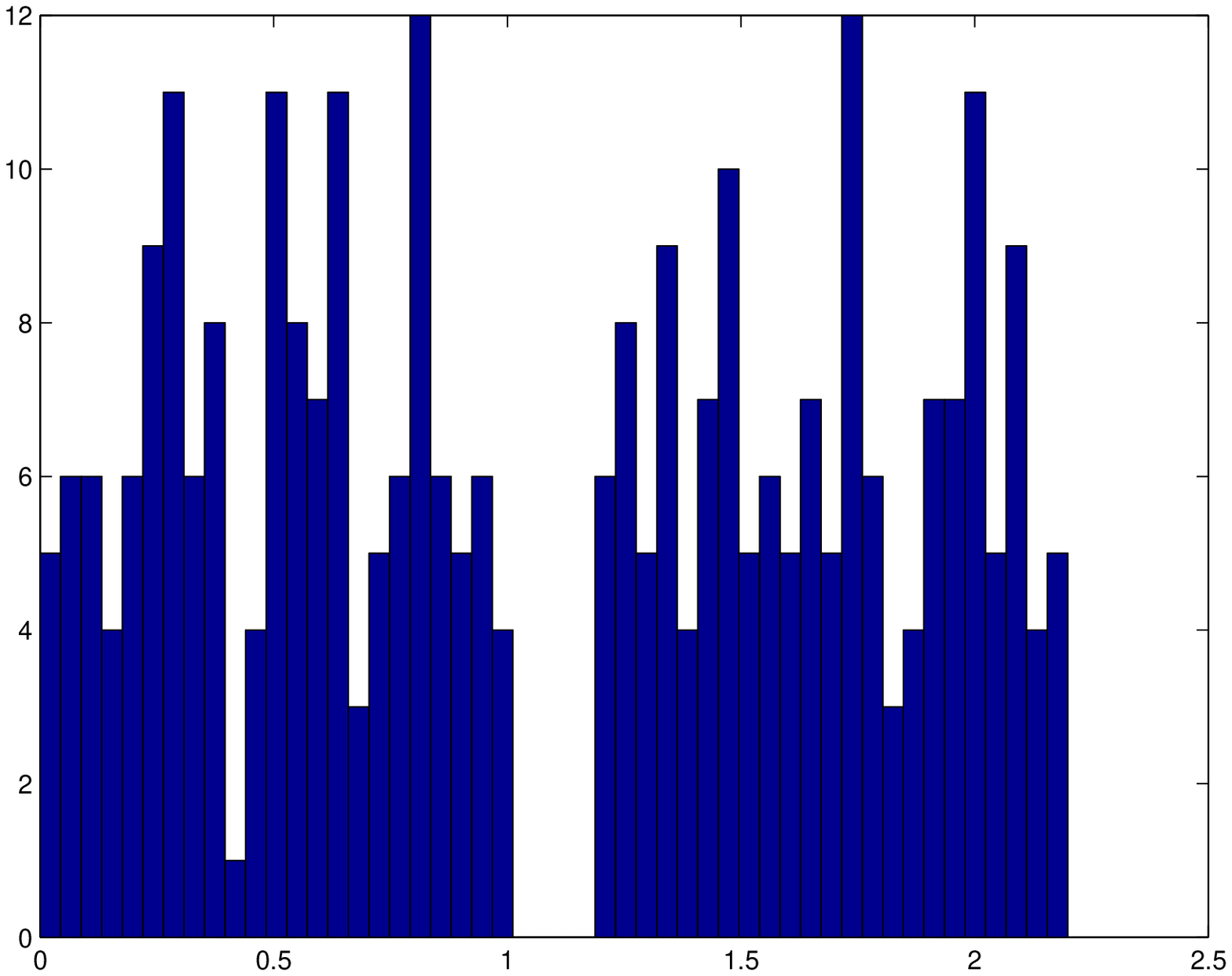,width=7.0cm,height=6.0cm}     
        \hspace{0.5cm}
     } \= 
     \subfigure[]
     {
        \psfig{figure=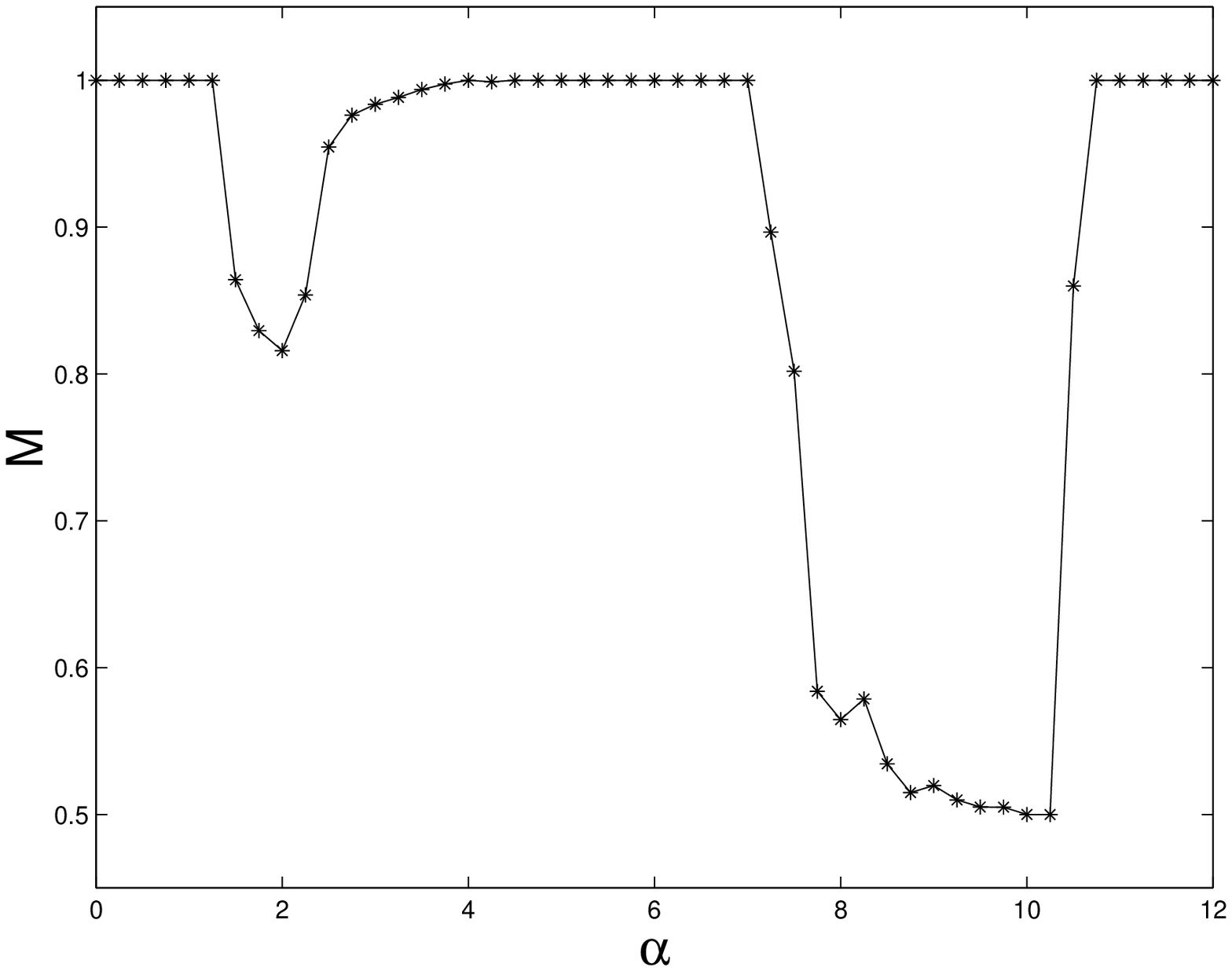,width=7.0cm,height=6.0cm} 
     } \\
     \centerline{
     \subfigure[]
     {
        \psfig{figure=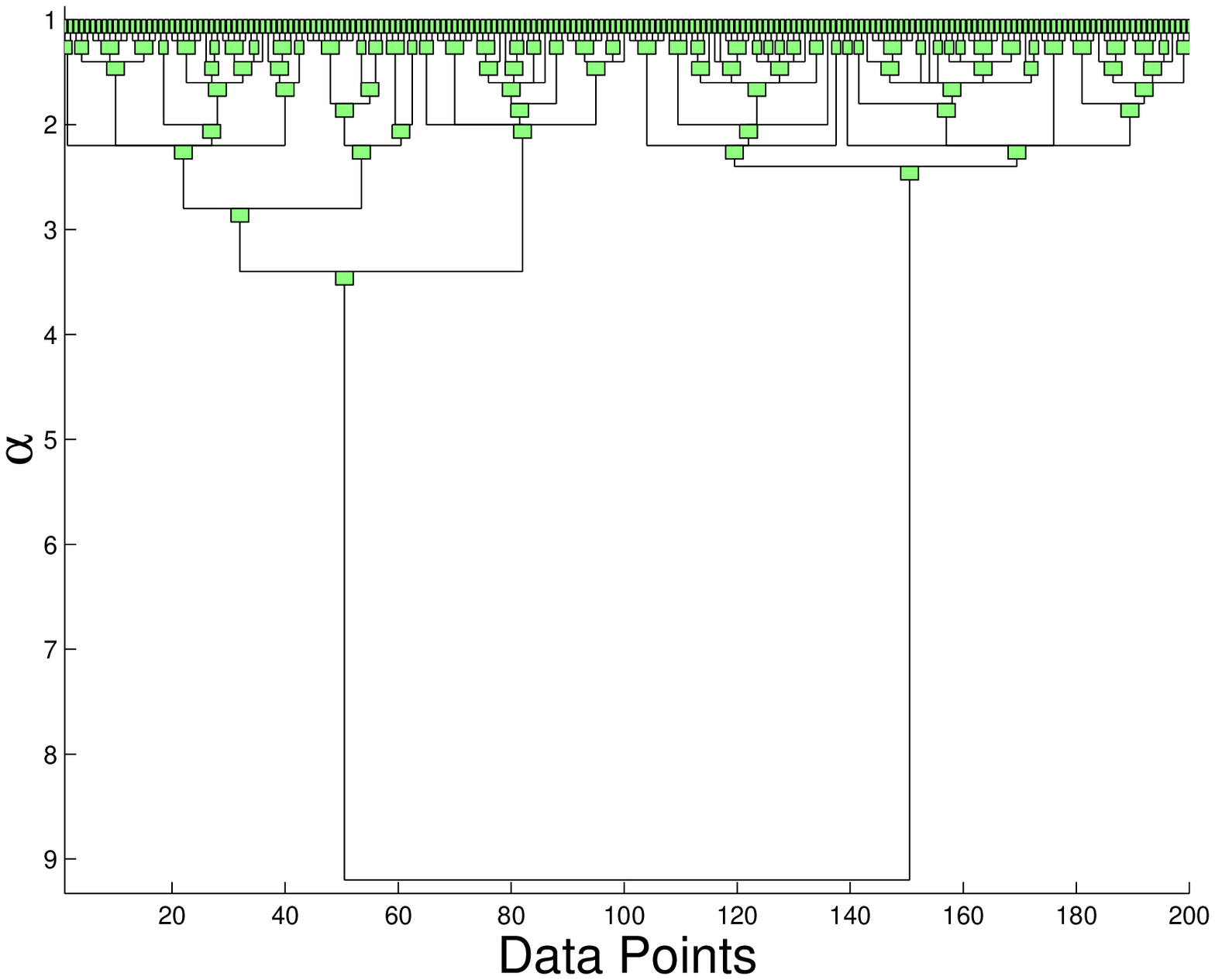,width=7.0cm,height=6.0cm} 
     }}
  \end{tabbing} 
    \caption{ Resampling results for a one-dimensional data set.	
        200 points were chosen uniformly from two clusters, 
	separated by $\Delta \lambda = 10$. Histogram of the data
	is given in (a). 
	We performed 100 resamples of 130 points, (\ie $f \approx 2/3$)
	to calculate $\M$.
	In (b) we plot
	$\M$ as a function of the resolution parameter $\alpha$. 
	The peak between $\alpha \approx 4$ and $\alpha \approx 7$
	corresponds to the correct two cluster solution, as
	can be seen from the dendrogram shown in (c). }
    \label{fig:data1d}
\end{figure}

\sction{Analysis of a One Dimensional Model}
To demonstrate the procedure outlined above we consider a
clustering problem, which is simple enough to allow an approximate analytical
calculation of the figure of merit $\M$ and its dependence on a parameter
that controls the resolution. 
Consider a one dimensional data set which consists of points $x_i, i-1,...,N$,
selected from two identical but displaced uniform distributions, such as the 
one shown in Fig. \ref{fig:data1d}.
The distributions are characterized
by the mean distance between neighboring points, $d=1/\lambda$,
and the distance or gap between the two distributions,  $\Delta$.
Distances between neighboring points within a cluster
are distributed according to the Poisson distribution, 
\be
P(s)=\lambda e^{-\lambda s} ds.
\label{eq:Poisson}
\ee  
The results of a clustering algorithm that reflects the underlying distribution
from which the data were selected
should identify two clusters in this data set. 

Consider a simple nearest-neighbor clustering algorithm, which
assigns two neighboring points to the same cluster if and only if
the distance between the two is smaller than a threshold $a$. 
Clearly, $\alpha = \lambda a$ is the 
dimensionless parameter of the algorithm that controls 
the resolution of the clustering.
For very small $\alpha << 1$ no two points belong to the same cluster, 
and the number
of clusters equals $N$, the number of points; 
at the other extreme,  $\alpha >> 1 $, 
all pairs of neighbors are assigned to the same cluster, and hence
all points reside in one cluster. 
Starting from $\alpha >> 1$ and reducing it gradually, one generates a
dendrogram. At an intermediate value of $\alpha$ we may get any number of
clusters between one and $N$. Hence, 
if we picked some particular value of 
$\alpha$ at which we obtained some clusters,
we must face the dilemma of deciding whether 
these clusters are ``natural", or are the result of the fluctuations 
(i.e. noise) present in the data.
In other words, 
we would like to validate the clustering solution obtained for the full data
set for a given value of $\alpha$. We do this  using 
the resampling scheme described above.

A resample is generated by  independently deciding for each
data point whether it is kept in the resample (with probability $f$),
or is discarded (with probability $1-f$). This procedure is repeated $m$ times,
yielding $m$ resamples.
All length scales of the original problem get rescaled by the
resampling procedure by a factor of $1/f$; the mean
distance between neighboring points in the resampled set 
is $d'=1/\lambda f$, and the distance between the two uniform distributions is
$\Delta'= \Delta / f$.
Clustering is therefore performed with a rescaled 
threshold $a'=a/f$ on any resample; the resolution parameter 
keeps its original value, $\alpha'=a'/d'=\alpha$. 

We first wish to get an approximate analytical expression for the
figure of merit $\M (\alpha)$ described above. To do this we consider
the gaps between data points, rather then the points themselves.

Let us denote by $b$ the distance between the data point $i$ (of the
original sample) and its nearest left neighbor: $b=x_i-x_{i-1}$,
with the two points on the same side of the gap $\Delta$.
We first assume that this edge is {\em not} broken by the clustering
algorithm, $b<a$. Given a resample that includes point $i$, we
define $b'$ in the same fashion.
The new, resampled left neighbor of $i$ resides in the same
cluster as $i$ if $b'<a'$; the probability that this happens is given by
\cite{Levine00}
\be
P_{1}(\beta)=\sum_{m=1}^{\infty}{ \frac{f^2 (1-f)^{m-1}}{(m-1)!}
  \gamma(m,\alpha / f - \beta)} .
\label{eq:P1}
\ee
where the dimensionless variable $\beta = \lambda b$ was introduced.
Here $\gamma(n,z)$ is Euler's incomplete Gamma function, 
\be
\gamma(n,z)=\int_{0}^{z}{e^{-t}t^{n-1}dt} ,
\ee
except that in our convention, we take $\gamma(n,z<0) = \gamma(n,0)$.

Similarly, if points $i$ and $i-1$ were {\it not} assigned to the same
cluster in
the original sample, then the probability that the same would happen
in a resample is\cite {Levine00} 
\be 
P_{2}(\beta)=\sum_{m=1}^{\infty}{ \frac{f^2 (1-f)^{m-1}}{(m-1)!}
  \Gamma(m,\alpha / f - \beta)} ,
\label{eq:P2}
\ee
where  $\Gamma(n,z)$ is the other incomplete Gamma function, 
\be
\Gamma(n,z)=\int_{z}^{\infty}{e^{-t}t^{n-1}dt} , 
\ee
and we take $\Gamma(n,z<0)=\Gamma(n,0)=(n-1)!$, 
so $B_m=1$ for $\alpha \leq f \beta$. 

We now calculate the index $\M$ in an approximate manner, by averaging $P_1$ and
$P_2$ over all edges. This calculation matches the
definition (\ref{eq:M}) of $\M$ in spirit.
For pairs residing within a true cluster, the averaging is done by 
integrating over all possible values of $b$;
\be
A(\alpha) = 
\int_{0}^{\alpha}e^{-\beta}{P_{1}(\beta)\; d\beta} + \int_{\alpha}^{\infty}e^{-\beta}{P_{2}(\beta)\; d\beta} \;.
\ee
For pairs which lie of different sides of the gap, 
we should only compare $a$ with the
size of the gap $\Delta$ :
\be 
B_{\delta}(\alpha) = \left\{
  \begin{array}{ll}
    P_{1(\delta)} & \alpha \geq \delta \\
    P_{2(\delta)} & \alpha < \delta
  \end{array} \right. , 
\ee
where the dimensionless variable $\delta = \lambda \Delta$ was introduced.

Clearly, in the one-dimensional example there
are much fewer edges of the second kind than of the first. 
This, however, is
not the case for data in  higher dimensions, so we give equal weights to
the two terms $A$ and $B$\footnote{The ratio between the two terms is of 
the order $(d-1)/d$, where $d$ is the dimensionality of the problem.
For high dimensional problems, this ratio is close to 1.}

\be
\M(\alpha) = \frac{1}{2}\left[A(\alpha) + B_{\delta}(\alpha)\right] .
\label{eq:anam}
\ee

\begin{figure}  
  \centerline{
    \psfig{figure=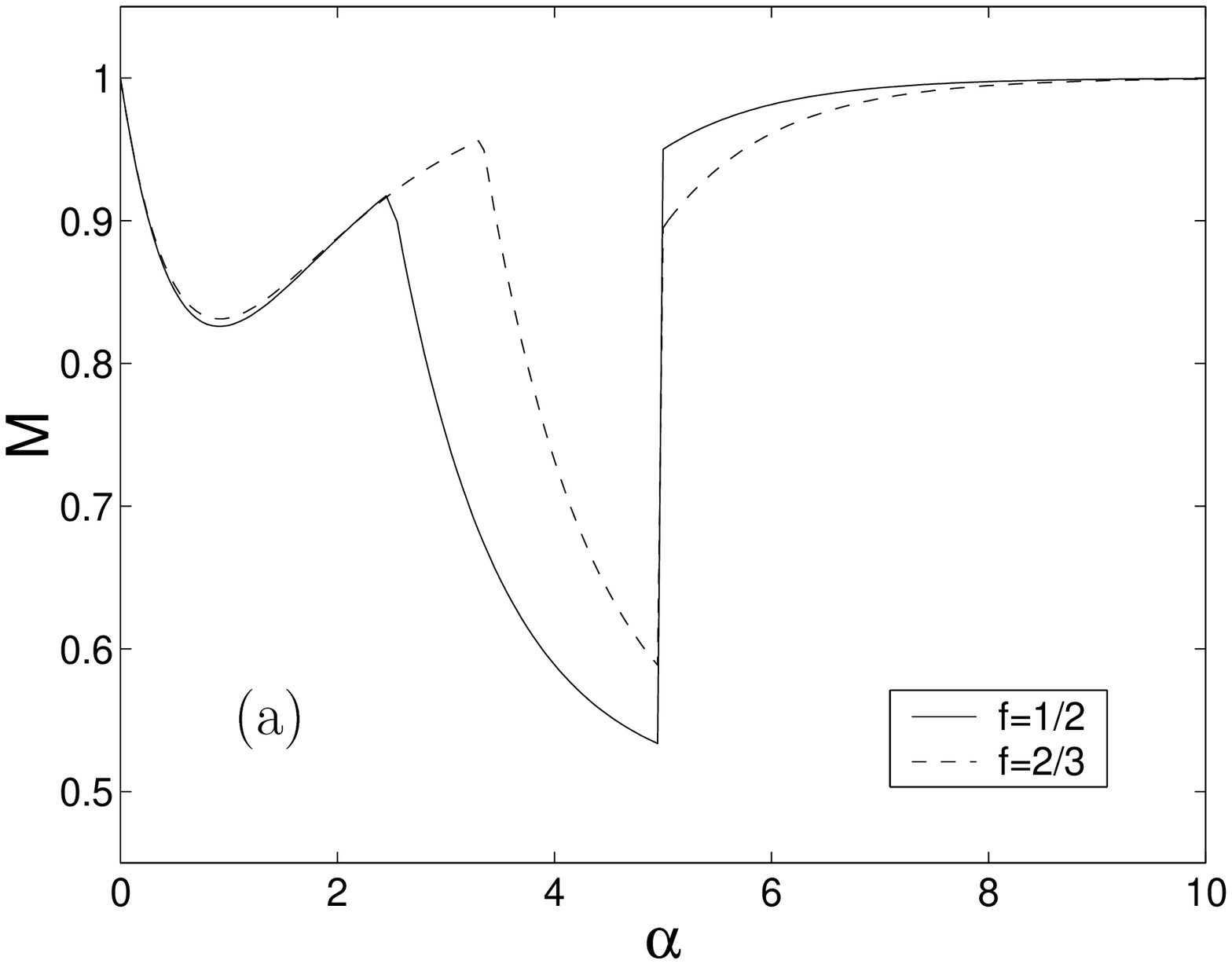,width=6.0cm} \hspace{1.0cm}
    \psfig{figure=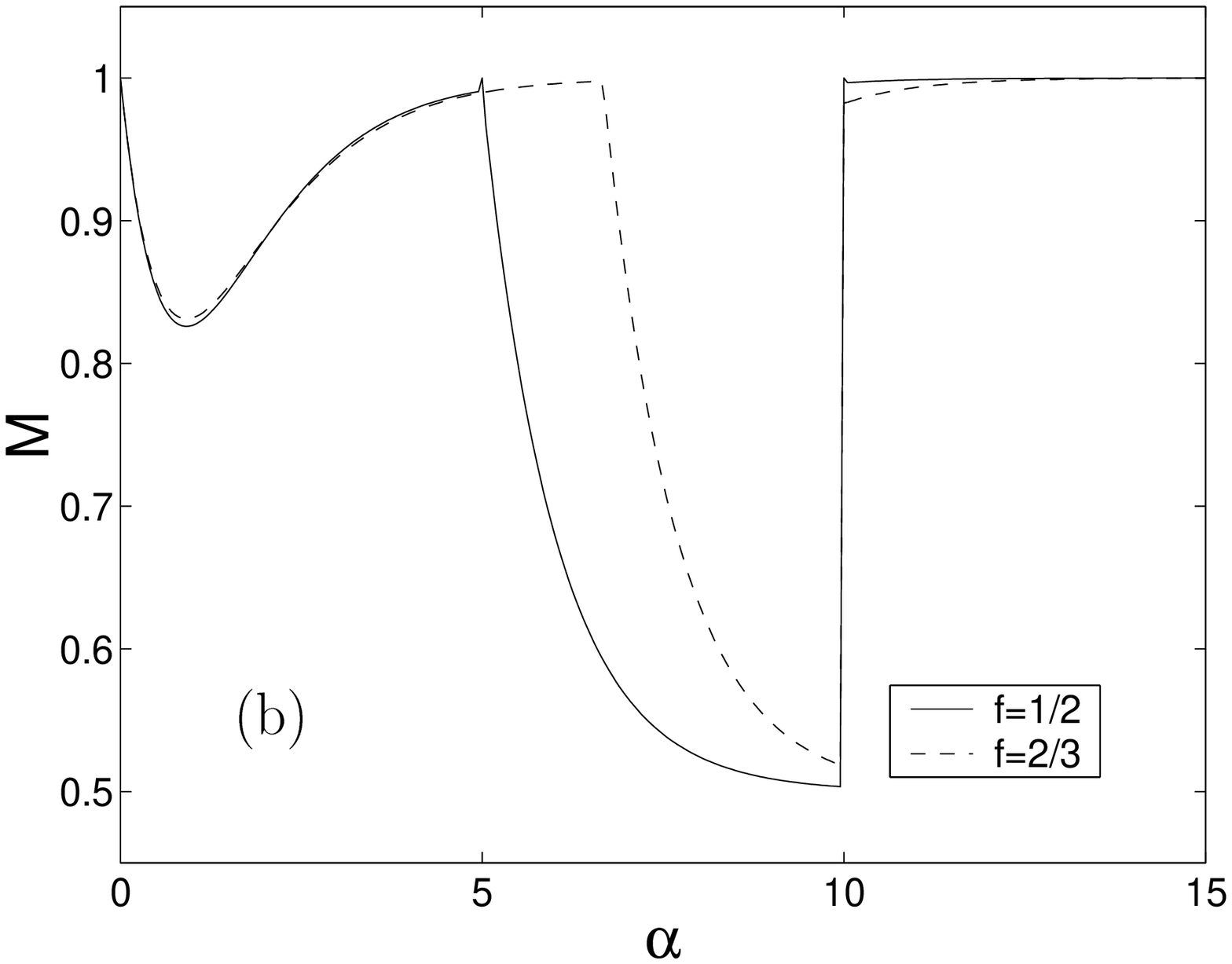,width=6.0cm}
    }
  \caption{Mean behavior of $\M$ as a function of the
    geometric threshold, 
    according to equation (\ref{eq:anam}), for
    two clusters. The function 
    is evaluated for
    the inter-cluster distances (a) $\Delta = 5 \lambda$ and
    (b) $\Delta = 10 \lambda$, with dilution parameters
    $f=1/2$ and $f=2/3$.}
  \label{fig:anaM}
\end{figure}

We now plot $\M$ as a function of the resolution parameter $\alpha$
for both $f=1/2$ and $f=2/3$ (figure \ref{fig:anaM}(a)),
assuming the inter-cluster distance $\delta = 5$.
A clear peak can be observed in both curves at $\alpha \simeq 2.5$ 
and $\alpha \simeq 3.3$, respectively. 
Similarly, for $\delta = 10$ clear peaks are identified at $\alpha \simeq 5$
and $\alpha \simeq 7$ (figure \ref{fig:anaM}(b)). As we will see,
these peaks correspond to the most stable clustering solution, which indeed
recovers the original clusters.
The trivial solutions, of a single cluster (of all data points) and the 
opposite limit of $N$
single-point clusters, are also stable, and appear as
the maxima of $\M(\alpha)$ at $\alpha << 1$ and $\alpha >> 1$, respectively.

In order to test how good is our analytic approximate evaluation of $\M$, 
we clustered a
one-dimensional data set 
of figure \ref{fig:data1d}(a), and calculated the
index $\M$ as defined in eq. (\ref{eq:M}). The data set consists of
$N=200$ data points sampled from two uniform distribution of mean
nearest neighbor distance $1/\lambda=1$ and shifted by $\Delta=10$. 
The dendrogram obtained by varying $\alpha$ is shown in 
Fig. \ref{fig:data1d}(c). It clearly exhibits two stable clusters;
stability is indicated by the wide range of values of $\alpha$ over which the
two clusters ``survive". Next, we generated
100
resamples of size 130 (\ie $f \approx 2/3$), and applied the
geometrical clustering procedure described above to each resample. 

By averaging over the different resamples
the figure of merit $\M$ was calculated for different values of the
dimensionless resolution parameter $\alpha$, as shown in figure
\ref{fig:data1d}(b). The peak between $\alpha \approx 4$ and $\alpha \approx
7$, corresponding to the most stable, ``correct" clustering solution, is clearly
identified. The agreement between our approximate analytical curve of figure
\ref{fig:anaM}(b) for $M(\alpha)$ and the numerically obtained exact 
curve of figure \ref{fig:data1d}(b)
is excellent and most gratifying.

\sction{Applications}


\subsection{Two Dimensional Toy Data}

\begin{figure}
  \centerline { \psfig{figure=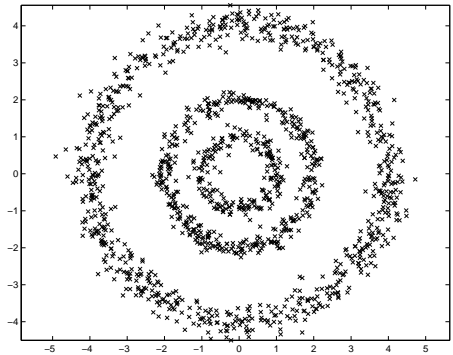,width=6.0cm} }
  \caption{ The three-ring problem. 1400 points are sampled from a three
    a three-component distribution described in the text. }
  \label{fig:circ}
\end{figure}

The analysis of the previous section predicts a typical behavior of $\M$
as a function of the parameters that control resolution; in particular, it
suggests that one can
identify a 
stable, ``natural" partition as the one obtained at 
a local maximum of the function $\M$. This prediction was based on an
approximate analytical treatment and backed up by numerical simulations of
one-dimensional data. Here
we demonstrate that 
this behavior is also observed for a
toy problem which consists of the
two-dimensional data set shown in figure \ref{fig:circ}.
The angular coordinates of the data points are selected from a 
uniform distribution, $\theta \sim \mbox{U}[0,2 \pi]$.
The radial coordinates are  normal distributed, $r \sim \mbox{N}[R,\sigma]$ 
around three different radii $R$. The outer ``ring'' ($R=4.0$, $\sigma = 0.2$)
consists of 800 points, and the inner ``rings'' ($R=2.0,1.0$, $\sigma = 0.1$)
consists of 400 and 200 points, respectively.

The algorithm we choose to work with is 
the {\em super-paramagnetic clustering} (SPC) algorithm, recently
introduced by Blatt \al \cite{Blatt96,Domany99}.
This algorithm provides a hierarchical clustering solution.
A single parameter $T$, called ``temperature", controls the resolution :
higher temperatures corresponds to higher resolutions. Variation of $T$
generates a dendrogram. The outcome of the algorithm depends also on an
additional parameter $K$, described below in subsection 4.3.1. The data of Fig.
\ref{fig:circ} were clustered with $K=20$.

\begin{figure}  
\centerline{\psfig{figure=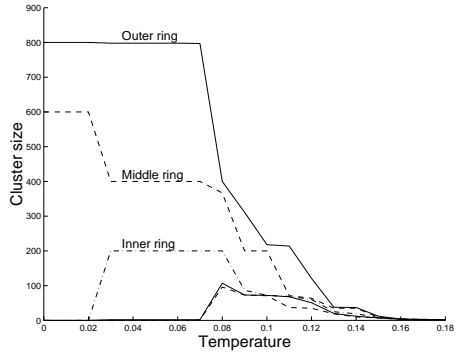,width=6.0cm}}
\caption{Clustering solution of the three ring problem as a function 
of resolution parameter (the temperature).}
\label{fig:cfull}       
\end{figure}

The results of the clustering procedure are presented in Figure \ref{fig:cfull}.
A stable phase, in which the three rings are clearly identified appears
at  the temperature range $0.3 \leq T \leq 0.8$.

In order to identify the value of $T$ that yields the
``correct" solution, we generated and clustered 20 different resamples 
from this toy data set, with a dilution factor of $f=2/3$. 
The resolution parameter (temperature) of each resample was rescaled
so that the transition temperature at which the single large cluster
breaks up agrees
with the temperature of the same transition in the original sample.

\begin{figure}[htbp] 
  \centerline{\psfig{figure=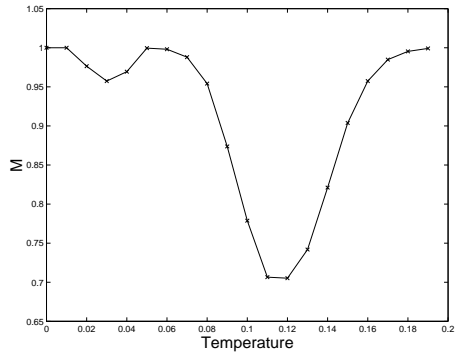,width=6.0cm}}
  \caption{$\M$ as a function of the temperature for the three-ring problem.}
  \label{fig:circres}
\end{figure}
The function $\M(T)$, plotted in figure \ref{fig:circres}, exhibits
precisely the expected 
behavior, with two 
trivial maxima and  an additional one at $T=0.05$. 
This value indeed  corresponds to the ``correct''
solution, of three clusters.

\subsection{A Single Cluster -- Dealing With Cluster Tendency}

A frequent problem of clustering methods is the so called 'cluster tendency';
that is, the tendency of the algorithm to partition any data, even when 
no natural clusters
exist. In particular, agglomerative algorithms, which provide a hierarchy
of clusters, always generate some hierarchy as a control parameter is varied. 
We expect this
hierarchy to be very sensitive to noise, and thus unstable against resampling.

We tested this assumption for the data set of figure \ref{fig:example}(a).
The test was performed using two clustering methods: the SPC algorithm mentioned
above, and the {\em Average-Linkage} clustering algorithm. The latter is an
agglomerative hierarchical method. It starts with $N$ distinct clusters, 
one for each point, and forms the hierarchy by successively merging the 
closest pair of clusters, and redefining the distance between all other 
clusters and the new one.
This step is repeated $N-1$ times, until only a single element remains.
The output of this hierarchical method is a dendrogram.
For more details see \cite{Jain88}.

\begin{figure}  
  \centerline{
    \psfig{figure=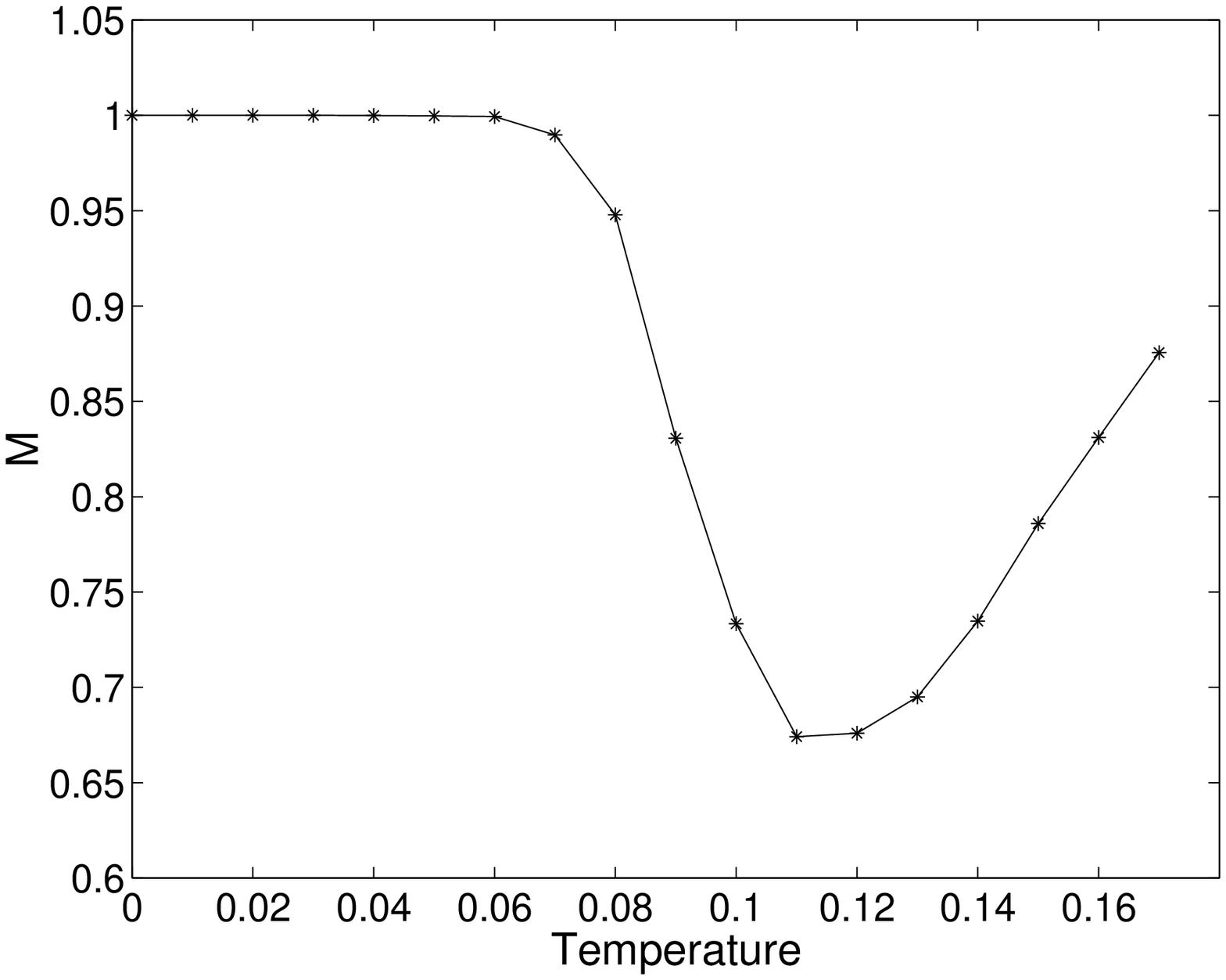,width=6.0cm}  \hspace{1.0cm}
     \psfig{figure=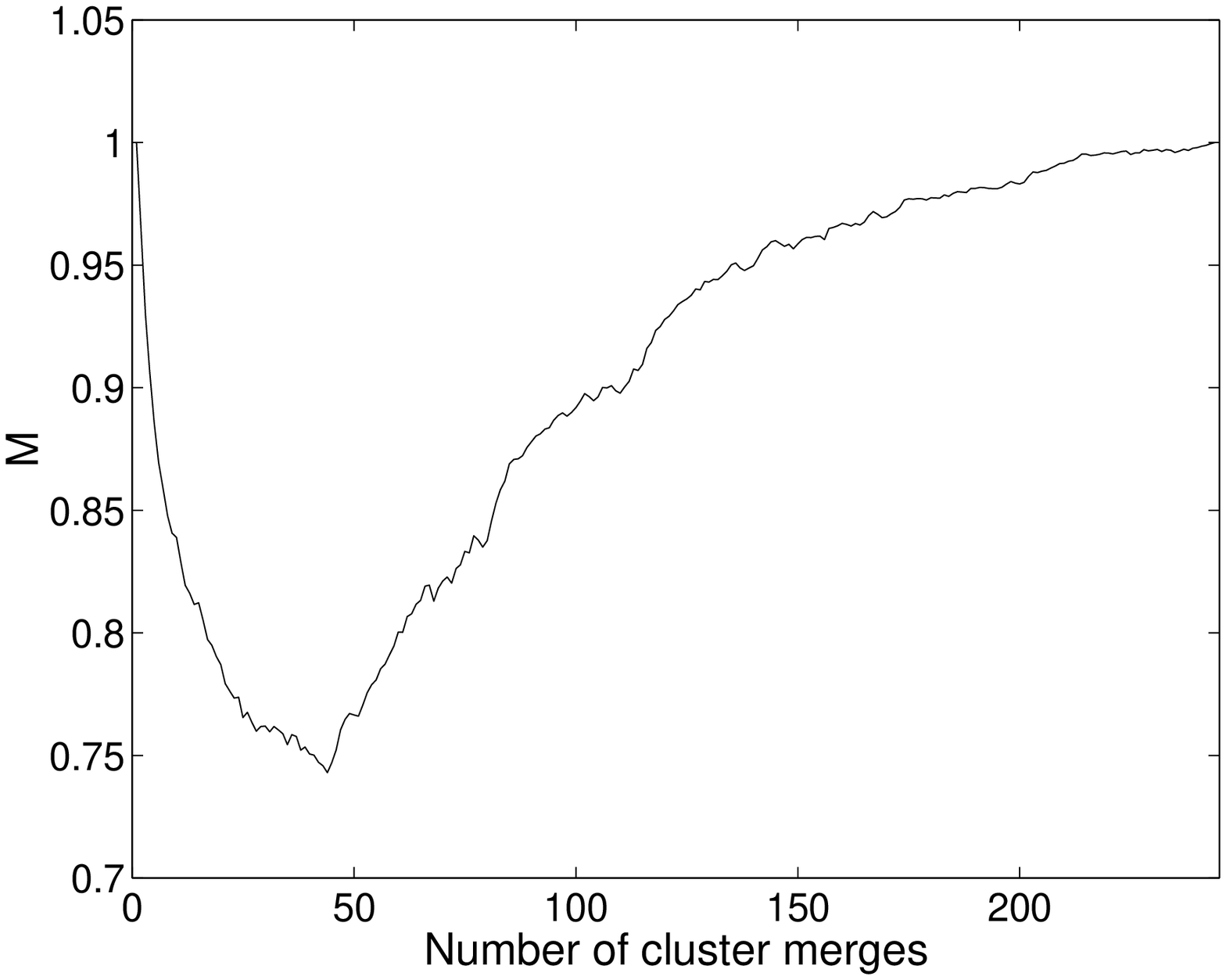,width=6.0cm}
    }
  \caption{$\M$ as a function of the resolution parameter for the single-cluster
        problem. Clustering is performed using the SPC algorithm (left) and the
        Average-Linkage algorithm (right).}
  \label{fig:tendency}
\end{figure}

We performed our resampling scheme with $m=20$ resamples and dilution factor
of $f=2/3$, for different levels of resolution. The results are shown in 
figure \ref{fig:tendency}. The only stable solutions 
identified by both algorithms are  the trivial ones, 
of either a single cluster or $N$ clusters. In this
case, obviously, the single cluster solution is also the natural
one.  

\subsection{Clustering DNA Microarray Data}


We turn now to apply our procedure to real-life data. We present here 
validation of cluster analysis performed for DNA Microarray Data.

Gene-array technology provides a broad picture of the state of a cell
by monitoring the expression levels of thousands of genes simultaneously.
In a typical experiment simultaneous expression levels of thousands of
genes are viewed over a few tens of cell cultures at different conditions. 
For details see \cite{Chee96,Brown99}.

The experiment whose analysis is presented here is on colon 
cancer tissues\cite{Alon99}.
Expression levels of $n_g=2000$ genes were measured for 
$n_t=62$ different tissues, 
out of which 40 were tumor tissues and 22 were normal ones.
Clustering analysis of such data has two aims:
\vspace{-0.3cm}
\begin{itemize}
\item[(a)] Searching for groups of tissues with similar gene
expression profiles. Such groups may correspond to normal vs tumor tissues.
For this analysis the $n_t$ tissues are considered as the data points,
embedded in an $n_g$ 
dimensional space. 
\item[(b)] Searching for groups of genes with correlated behavior.
For this analysis we view the genes as the data 
points, embedded  in an $n_t$ dimensional space and we hope to find
groups of genes that are part of the same 
biological mechanism.
\end{itemize}
Following Alon \al, we normalize each data point (in both cases) such that
the standard deviation of its components is one and its mean vanishes. This
way the Euclidean distance between two data points is trivially related to the 
Pearson correlation between the two.

\subsubsection{Clustering tissues}
The main purpose of this analysis is to check whether
one can distinguish between tumor and normal tissues 
on the basis of gene expression data. 
Since we know in what aspect of the data's structure we are interested,
in this problem the working resolution
is determined as the value at which  
the data first breaks into two (or more) large clusters (containing, say,
more than ten tissues).

There may be, however, other parameters for the clustering algorithm, which
should be determined. For example, the SPC algorithm has a single parameter
$K$, which determines how many data points are considered as neighbors
of a given point. The algorithm places edges or arcs that connect pairs
of neighbors $i,j$, and assigns a weight $J_{ij}$ to each edge, whose value
decreases with the distance $\vert {\vec x}_i - \vec{x}_j \vert$
between the neighboring data-points.
Hence the outcome of the clustering process may depend on $K$, the number of
neighbors connected to each data point by an edge. 
We would like to use our resampling method to determine
the optimal value for this parameter.

\begin{figure}
     \centerline{
        \psfig{figure=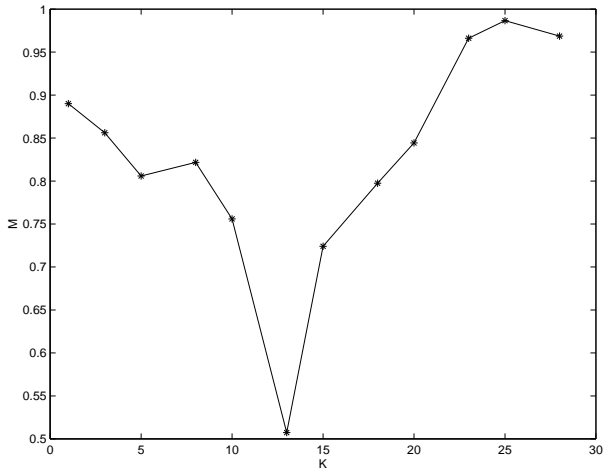,width=8.0cm}
     } 
    \caption{Figure of merit $\M$ as a function of $K$ for the colon tissue data.}
    \label{fig:col_val_M}
\end{figure}

We clustered the tissues, using SPC, for several values of $K$. For each case
we identified the temperature at which the first split to large clusters
occurred. For each case we performed the same resampling scheme, 
with $m=20$ resamples
of size $\frac{2}{3}n_t$, and calculated the figure of merit $\M (K)$. 
The results obtained for several values of $K$ are
 plotted in figure \ref{fig:col_val_M}. 
 Very low and very high values of $K$ give
similarly stable solutions. In the low-$K$ case each point is practically 
isolated,
and the data breaks immediately into microscopic clusters. 
The high-$K$ case is just
the opposite -- each points is considered ``close'' to very many other points, 
and no macroscopic partition can be obtained.  

\begin{figure}
     \centerline{
          \psfig{figure=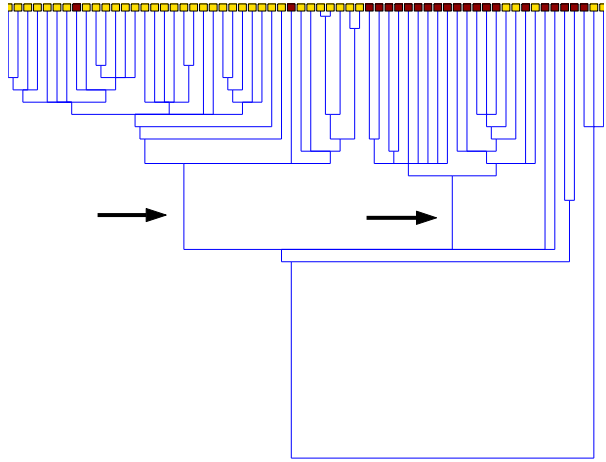,width=8.0cm}
     } 
    \caption{Clustering of the tissues, using K=8. Leaves of the tree
are colored according to the known classification to normal (dark) 
and tumor (light). The two large clusters marked by arrows recover the 
tumor/normal classification.}
    \label{fig:tndend}
\end{figure}

At $K=8$ we observe, however, another peak in $\M$. At this $K$
the clustering algorithm yields indeed two large clusters, which correspond to the two
``correct" clusters of normal and tumor tissues,
as shown on Fig. \ref{fig:tndend}. Such solutions appear also 
for some higher values of $K$, but 
in these cases the clusters are  not stable against resampling.

\subsubsection{Clustering genes}

Cluster analysis of the genes is performed in order to identify
groups of genes which act cooperatively. Having identified a
cluster of genes, one may look for a biological mechanism that
makes their expression
correlated. Trying to answer this question,
one may, for example, identify common promoters of these genes \cite{Getz99a}.
One may also use one or more clusters of genes to reclassify the tissues
\cite{Getz00}, looking for clinical manifestations associated with the expression
levels of the selected groups of genes. 
Therefore in this case it is important to  assess
the reliability of each particular gene cluster separately.

\begin{figure}
     \centerline{
        \psfig{figure=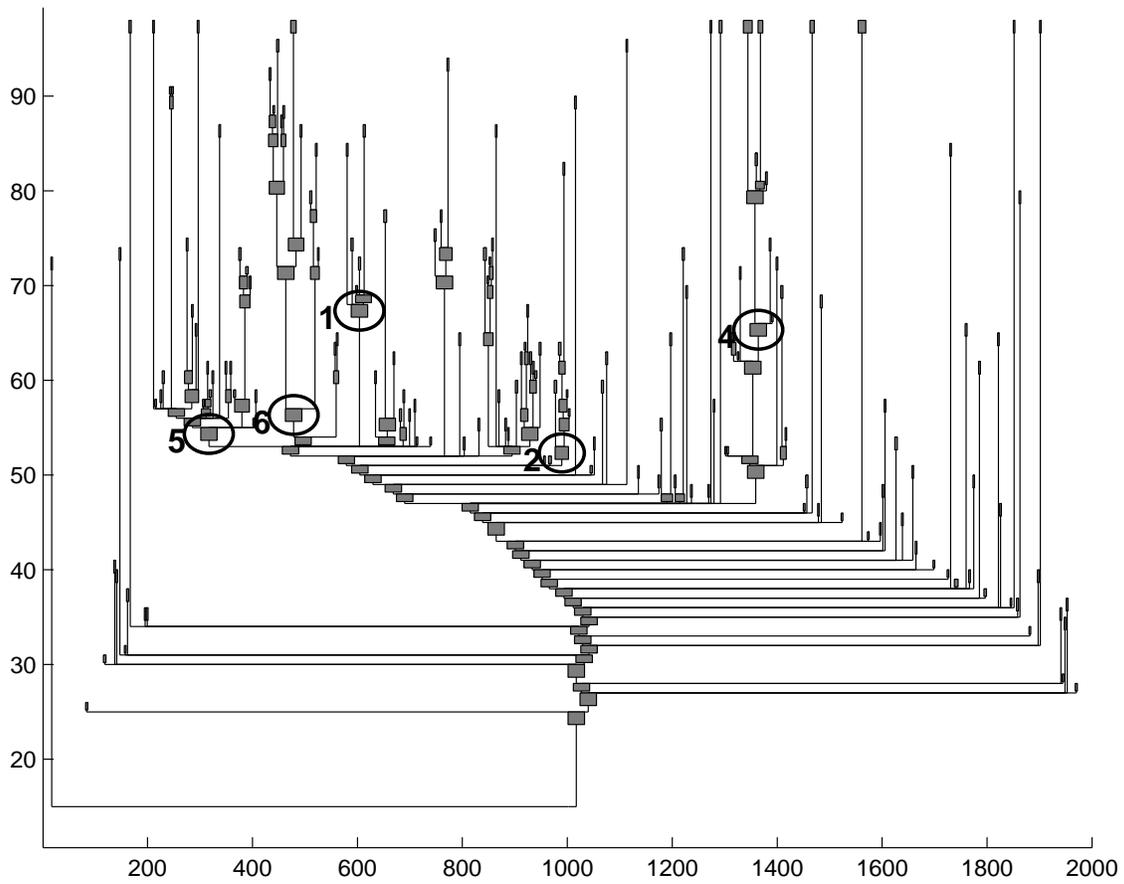,width=15.0cm}
     } 
    \caption{Dendrogram of genes. 
        Clusters of size 7 or larger are shown as boxes.
        Selected clusters are circled and numbered, as explained
        in the text.}
    \label{fig:colspcdend}
\end{figure}

The SPC clustering algorithm was used to cluster the genes. A resulting
dendrogram is presented in Fig. \ref{fig:colspcdend}, in which each box
represents a cluster of genes. 
Our resampling scheme
has been applied to this data, using $m=25$ resamples of size 1200 
( $f = 0.6$).
This time, however, we calculated $\M (C)$, {\it for each cluster C} 
of the full original
sample, separately. That is, we calculated $\M$ only for the
points of a single cluster $C$, at the temperature at which it was identified,
and then moved on to the next cluster.
We therefore get a stability measure for each cluster.
Next, we  focus our attention on clusters of the highest scores.
First, we considered the top 20 clusters. If one of these clusters
is a descendent of another one from the list of 20,
we discard it. After this pruning we were
left with 6 clusters, which are circled and numbered in 
figure \ref{fig:colspcdend} (the numbers are not related to 
the stability score).

We are now ready to interpret these stable clusters. The first three 
consist of known families of genes.
\#1 is a cluster Ribosomal proteins. The genes of cluster \#2
are all Cytochrome C genes, which are involved in energy transfer.
Most of the genes of cluster \#3 belong to the HLA-2 family, which are 
histocompatability antigens. 

Cluster \#4 contains a variety of genes, some of which
are related to metabolism. When trying to cluster the tissues
based on these genes alone, we find a stable partition to two
clusters, which is not consistent with the tumor/normal labeling.
At this point, we are still unable to explain this partition.

Clusters \#5 and \#6 also contain genes of various types. The genes
of these clusters have most typical behavior: all the genes
of cluster \#5 are highly expressed in the normal tissues but
not in the tumor ones; And all genes of cluster \#6 are
the other way around. 

To summarize, clustering stability score based on resampling enabled
us to zero in on clusters with typical behavior, which may have biological
meaning. Using resampling enabled us to select clusters without
making any new assumption, which is a major advantage in exploratory 
research. The downside of this method, however, is it's computational
burden. In this experiment we had to perform clustering analysis
20 times for a rather large data set. This would be the typical case
for DNA microarray data.

\sction{Discussion}
This work proposes a method to validate clustering analysis results, based
on resampling. It is assumed that a cluster which is robust to resampling
is less likely to be the result of a sample artifact or fluctuations.

The strength of this method is that it requires no additional assumptions.
Specifically, no assumption is made either about the structure of the
data, the expected clusters, or the noise in the data. No information,
except the available data itself, is used.

We introduced a figure of merit, $\M$, which reflects the stability of the cluster
partition against resampling. The typical behavior of this
figure of merit 
as a function of the resolution parameter allows
clear identification of natural resolution scales in the problem.

The question of  natural resolution levels is inherent to the 
clustering problem, and thus emerges in any clustering scheme.
The resampling method introduced here is general, and applicable
to any kind of data set, and to any clustering algorithm. 

For a simple one-dimensional model we derived an analytical 
expression for our figure of merit and its 
behavior as a function of the resolution
parameter. Local maxima were identified for values of the 
parameter corresponding to stable clustering solutions. Such
solutions can either be trivial (at very low and very high
resolution); or non-trivial, revealing genuine internal structure
of the data.

Resampling is a viable method provided
the original data set is large enough, so that a typical resample
still reflects the same underlying structure. If this is the case,
our experience shows that a dilution factor of $f \simeq 2/3$ works well
for both small and large data sets.

\subsection*{Acknowledgments}

This research was partially supported by the Germany - Israel Science Foundation
(GIF). We thank Ido Kanter and Gad Getz for discussions.

\renewcommand{\baselinestretch}{0}

\end{document}